\documentstyle[pre,preprint,aps]{revtex}
\begin{document}
\draft
\title{A Two Scale Model for Aggregation and Etching}
\author{George C. John}
\address{Physics Department, Indian Institute of Technology, Kanpur,
U.P.-208016, INDIA}
\author{Vijay A. Singh\thanks{on leave from I.I.T. Kanpur}}
\address{Solid State Electronics Group, Tata Institute of
Fundamental Research, Bombay-400 005, INDIA}
\date{ }

\maketitle

\newcommand{\etal}{{\em et al.$\,$}}
\newcommand{\dw}{\mbox{$\Delta W \,$}}
\newcommand{\hr}{\mbox{$\dot{h} \,$}}

\bibliographystyle{prsty}
\begin{abstract}
	We propose a dual scale drift-diffusion model for interfacial
growth and  etching processes. The  two scales  are: (i)  a depletion
layer  width \dw  surrounding the aggregate and (ii)  a drift  length
$l$. The interplay between  these  two  antithetical scales yields  a
variety   of  distinct   morphologies  reported  in   electrochemical
deposition of metals,  viscous  fingering in  fluids  and  in  porous
silicon  formation.   Further,  our  algorithm  interpolates between
existing  growth models  (diffusion  limited  aggregation,  ballistic
deposition and Eden) for limiting values of \dw and $l$.
\end{abstract}

\section{Introduction}
	The  study  of  complex  patterns  in   dimension  $d=2$  has
attracted a great  deal  of attention  in the past decade and a half.
Systematic studies of emergent patterns in electrochemical deposition
(ECD) of metals and of the  morphologies obtained during fluid-fluid
displacement   in   a   Hele-Shaw   set    up   have   been   carried
out\cite{benj90}.   On  the other hand,  complex  pore  geometries of
anodically etched silicon have also evoked considerable interest,  on
account of the technological promise of  porous silicon\cite{smit92}.
A simple model  which  would  describe the  fascinating  varieties of
morphologies obtained both  in aggregation and dissolution would be a
desirable objective.

	In electrochemical  deposition experiments,  two  parameters,
the electrolyte concentration  (C) and the applied  potential (V) are
tuned to obtain (i) dendritic structures,  both thick and  needlelike
(or  stringy);  (ii)  Dense  branching morphologies (DBM),  which are
homogeneous and (iii) randomly ramified self similar structures as in
diffusion   limited   aggregation(DLA)\cite{sawa86,grie86}.    In   a
Hele-Shaw  cell,  similar transitions are observed  when the pressure
and  surface tension  are  varied\cite{benj90}.   In silicon,  anodic
etching  gives rise to  differing pore morphologies  depending on the
anodization      potential     and      the      substrate     doping
level\cite{smit92,beal85}.

	A first principles explanation of such phenomena will have to
encompass (i)  the  diffusive  field, (ii) the  Laplace field,  (iii)
convective   processes  and   (iv)  surface  tension,  curvature  and
underlying anisotropy effects.  Hence, even  numerical  solutions may
prove difficult.  Simple formal approaches  have  been  attempted for
obtaining         stability         conditions         at         the
interface\cite{mull63,grie87,gari89}.  It has also been  hypothesized
that  the emergence of different  characterizing length scales is the
result  of  the  interplay between  the Laplace and diffusion  fields
governing  the  growth process  in  ECD\cite{gari89}.  Similar models
have  been  proposed  to  model  the  anodic etching  of  silicon  in
hydroflouric  acid.   A  phenomenological  Schottky  diode model  for
porous silicon formation\cite{beal85} invokes a depletion layer width
at  the interface with a  substantial  barrier  lowering at the  pore
tips.

	On the other hand, a  simple algorithmic approach has been
adopted in the last decade and a  half to successfully capture the
basic   patterns  mentioned above\cite{herr86,meak88,stan85}.  The
most   common amongst these  is the  diffusion limited aggregation
(DLA)\cite{witt81} which  results  in  a  self similar   structure
resembling   zinc  leaves grown in   an  electrolyte under certain
limiting   conditions.  Generalizations   of  the DLA   have  been
proposed   by the introduction  of additional  parameters into the
model.     Surface  tension effects  have     been modeled by  the
incorporation   of   a sticking    probability  at  the  aggregate
surface\cite{vics84}.  This resulted in  a transition from the low
density  fractal    clusters to   regular  patterns.   Recently, a
multiwalker DLA  model has been proposed\cite{erle93}  wherein two
parameters,  namely the particle concentration  and the width of a
``migrational envelope''   are  tuned to    obtain dense branching
morphologies as well as DLA patterns.

	In this article, we propose a single walker, two parameter
drift-diffusion model  to simulate growth   (and dissolution) in a
wide variety of systems.  Section 2 states the algorithm  employed
in the  growth process.  Section 3  summarizes  the various growth
patterns obtained in the simulations with differing parameters and
correlates   them to   the    morphological classes  reported   in
experimental literature. Section 4 constitutes a brief discussion.

\section{The drift-diffusion model}

	It will be a desirable goal to arrive at a model which can
reproduce the  various morphologies obtained from different growth
phenomena. Modeling the  transition from one  class of patterns to
another will also be  an  interesting objective. Such  transitions
can be represented as resulting from the  interplay of two or more
parameters \cite{shoc92a,shoc92b}. Different  morphologies  result
in the various limiting cases. However, to maintain the simplicity
of  such  an  approach, a  minimal   set of  parameters  should be
employed.

	In the classical DLA model\cite{witt81}, a particle starts
its random walk  at an infinitely  large distance away from a seed
or cluster.  Its random diffusive motion  is terminated the moment
it comes into contact with the aggregate cluster.  This simulation
models  diffusion in  the  low concentration  limit  and  does not
employ any control  parameters.  To control the diffusive process,
the introduction  of other  variables such  as a  diffusion length
\cite{smit88,smit89} or  particle concentration \cite{voss84} have
been suggested in literature.

	It  is    conceivable that very   close  to  the aggregate
surface,  the  particle movement  is  no longer  controlled by the
macroscopic    diffusion field,  but   by   microscopic, localized
phenomena. To   model such processes,  additional parameters  will
have to be introduced into the simulation.  We attempt to develop
an algorithm  which models pattern  formation  as a result  of the
interplay of a macroscopic field and localized surface phenomena.

	 The main control   parameters in this  model are:  (i)  a
depletion layer width \dw which  controls the diffusion and (ii) a
drift  length  $l$ governing  the field   driven processes in  the
proximity of  the   aggregate. The  design  of  the   algorithm is
outlined below:

	(i) The  particles  are launched, one at  a   time, from a
randomly  chosen lattice site beyond  the depletion layer boundary
at a  distance  \dw from the  surface  of the  aggregate. This  is
illustrated in Fig. 1.  We begin with a circular depletion zone of
radius \dw  around a central seed.    As the aggregate  grows, the
depletion   zone boundary  is  modified  to  follow the  aggregate
contour at a distance $\Delta W$ as shown in the figure.

	(ii)  The  particles execute random  walks  as in ordinary
off-lattice DLA algorithms \cite{meak88,ball85}.

	(iii) The  moment a   part  of the aggregate    surface is
encountered  within  a radius  $l$ of  the  particle location, the
random  walk  is terminated. The   particle is then   moved to the
surface (in a sense,  ``field  driven'') and  becomes part of  the
aggregate.

	The simulations were stopped when the  aggregate had grown to
a radius of around 200 units, one unit being the diameter of a single
particle. Approximately $10^4-10^5$ particles had to be launched for
various cases.

	The  two control  parameters employed   in this simulation
have also been suggested elsewhere in literature.  To model porous
silicon formation, a   ``finite diffusion length'' similar  to our
\dw has been suggested\cite{smit88}.  The drift  length $l$ in our
simulation is  in a way similar to  the width of the ``migrational
envelope''  of Erlebacher \etal \cite{erle93,erle94}.  A depletion
layer width $\Delta W$ and a drift diffusion  length $l$ have been
used  in conjunction with other  parameters  to successfully model
porous     silicon   formation   \cite{john95a}.    Our  algorithm
interpolates between  existing  growth models in  various limiting
cases.  In the limit $l=1$ and  $\Delta W \rightarrow \infty$, the
algorithm is identical to the DLA.  $\dw=0$ and $l=1$ approximates
the Eden  limit. As $l$ becomes  larger than $\Delta W$  (the case
depicted  in fig.  1), a  fraction of  the generated particles are
directly   transported    to   the  surface     without undergoing
diffusion. This is   akin  to ballistic deposition.    When  $l\gg
\Delta W$, the ballistic process dominates the diffusive process.
\section{Results}

	Fig.  2 depicts the  variety of morphologies obtained with
varying \dw  and  $l$.  For   small  $l/\dw$, we obtain   DLA like
patterns (Fig.  2a).  A simple mass-radius scaling calculation for
this pattern yielded a    fractal dimension of $1.65  \pm   0.05$.
Stringy structures similar  to those  reported in  electrochemical
deposition at high voltages\cite{sawa86} are seen in the large \dw
limit  with $l/\dw \simeq 1$ (Fig.   2b).  The mass-radius scaling
exponent fluctuates  around  $\sim 1.45$   for this pattern.   The
stringy morphology observed in some experiments is essentially one
dimensional  in character  \cite{grie86}.  This is  reproduced for
very large $\Delta W (=l) > 25$.  As $\Delta W (=l)$ is increased,
the scaling exponent  approaches   unity (see figure 4).    On the
other hand, for small \dw and $l/\dw  \simeq 1$, we obtained short
dense branches exhibiting a relatively  smooth front which remains
nearly circular throughout  the growth period  (Fig.2c).  This has
been identified in  experimental literature as homogeneous or  tip
splitting    patterns\cite{sawa86,grie86,grie87,benj86}.  The mass
radius  scaling  for  these structures    resulted  in a   fractal
dimension $\simeq 2$.  For $l/\dw \geq 3$, thick dendritic growths
with side branching was obtained (Fig.  2d).   For large $\Delta W
(>  5)$, the patterns  become too  inhomogeneous  to show any well
defined scaling \cite{sawa86}.  However, for small $\Delta W$, the
model approaches the  Eden  limit and the  resultant  patterns are
somewhat homogeneous.

	  The occurrence of  dendritic growth in such a simulation
is  surprising, since it is   well  understood that anisotropy  is
required  in  the  interfacial   dynamics for side   branching  to
occur\cite{benj90}.  The present  algorithm  admits an  underlying
anisotropy in the way a   particular surface site is chosen,  when
more than one point on the aggregate surface falls within a radius
$l$ of the random walker.  This selection  was done in three ways.
(i)  The nearest  site  was chosen.   (ii)   A site was chosen  at
random.  (iii) The site nearest  to the radial line connecting the
random walker and  the central seed  was  chosen. The morphologies
remained similar for the case $l\leq \Delta W$.  The three methods
yielded different results in  the  limit $l  \gg  \dw$.  Dendritic
growth was  observed only in case (iii),  which corresponds to the
imposition  of  a preferred     radial   field. In the    case  of
electrodeposition, this could be interpreted as the applied radial
electrostatic field.

	Figures  3 and 4  show   the variation in the  mass-radius
scaling  exponent $d_f$ with   varying  values of $\Delta W$   and
$l$. The values  obtained for various $\Delta W$  when $l$ is kept
constant ($l=2$) are plotted in Fig.3. $d_f$  is seen to vary from
2.00 to $1.65\pm0.05$  as $\Delta W$  goes from very small  ($\sim
1$)  to  large  ($  >  15$) values.   This  illustrates  a  smooth
transition between Eden type growth and DLA. As $\Delta W$ is kept
equal to $l$ and varied,  $d_f$ decreases steadily with increasing
$\Delta W$ ($=l$)   and  approaches unity.  This is  depicted   in
Fig. 4 and represents the emergence of stringy structures as shown
in Fig.2.

	Fig.      5   codifies      our    observations  into    a
``phase''\cite{phase} plot.  The plot was constructed on the basis
of approximately a hundred  patterns, grown to aggregate diameters
of around  400 units.  The  transitions from one phase  to another
being continuous,  the  boundaries  are not rigidly   defined. For
example,  in the $l \gg   \Delta W$ case,  the dendritic  patterns
observed at large $\Delta W$ become denser and more homogeneous as
$\Delta  W$  approaches  zero.   The  phases   depicted  have been
reported   earlier in   electrochemical deposition     experiments
\cite{sawa86,grie86} and related processes\cite{benj90}.

	Several    studies  of two  dimensional  growth   on a one
dimensional substrate have  been reported in  the past. Matsushita
\etal   have    grown  zinc  ``trees''      on a   linear   carbon
cathode\cite{mats85}.  Such processes have also been modeled using
a DLA like algorithm\cite{meak83}.  The  etching process in porous
silicon         may         also           be               viewed
similarly\cite{smit88,erle94,john95a}. This   is  achieved in  our
model by using a  linear substrate which is  represented by a side
of   the   lattice.  The deposition    is  no  longer radial,  but
``quasi-one dimensional''.

	Fig.  6 depicts the dependence of the mean  aggregate density
on the drift length $l$. Deposits were grown on a linear substrate of
length  300  units  for  a  given  \dw.   A  cross-over  behavior  is
discernible,  with  a minima at $\dw = l$.  This can be understood in
terms of our phase  plot (fig.  5).  The  minima  corresponds  to the
stringy  region, where $\dw \simeq l$. On either side of  the
stringy  phase, there  exists phases  of  higher density.  A  similar
behavior is  seen in  porous silicon where  transitions  are observed
from a uniform network  of  thin  pores to shorter,  wider  pores  on
increasing the  doping level in  the  substrate\cite{beal85}.  On the
other  hand, increasing the  applied potential leads to the formation
of thick and relatively linear pipe-like pores\cite{smit92}.

\section{Discussion}

	The finite diffusion length  model of Smith and co-workers
is  known  to  exhibit a  cross-over from  fractal  to non-fractal
clusters when the    aggregate size exceeds the diffusion   length
\cite{smit89,bowe91}.  In our    model,   a similar behavior    is
discernible for large  aggregate  sizes  when the  simulation   is
carried  out  for deposition  on   a  linear  substrate in  a  two
dimensional lattice (see Figure  7a).  The density decreases  with
depth  and finally  becomes  almost  depth  independent.  This  is
indicative of a non-fractal behavior (ie, in the depth independent
region).  However, in the case of radial deposition onto a central
seed, no such transition was observed for aggregate diameters from
100 to  1000  lattice units.   The   mass radius scaling  exponent
remained independent of the   aggregate size. This is depicted  in
figure  7b which  plots  the aggregate   mass against the  cluster
radius. The constant slope of  the log-log  plot is indicative  of
scale invariance over a large range of diameters (100-1000).

	One   can  explore  the  phenomenological  relevance  of  the
parameters, \dw and $l$. The  significant role  played by microscopic
dynamics  in  modifying  the  solutions of macroscopic  analyses  has
received growing appreciation  in the  recent past\cite{benj90}.   In
the diffusion field  versus drift  field dynamics governing growth in
this  simulation,  \dw is a  control  parameter  for the  macroscopic
diffusion field, whereas $l$ models microscopic surface  effects.  An
example of a  phenomenological model  employing  similar arguments is
the  Beale  model\cite{beal85}  for  porous silicon  formation  which
hypothesizes the existence of a depletion layer whose width varies as
\begin{equation}
    \dw \propto\left[\frac{(V_{BI}-V_A)}{n_0}\right]^{\frac{1}{2}}
\end{equation}
	where  $V_a$  is  the anodization potential,  $V_{BI}$  is  a
constant  built in voltage and  $n_0$  is the substrate doping level.
On the other hand, the parameter $l$  can be  related  to the barrier
lowering $(\Delta\phi_s)$ due to  microscopic surface irregularities.
This lowering, $\Delta  \phi_s  \propto \sqrt{E_s}$, $E_s$ being  the
enhanced local electric field.  For  planar interfaces,  $E_s \propto
\phi_s/\dw$  where $\phi_s$  is  the overall barrier height.   Due to
surface irregularities, this height is enhanced locally as
\begin{equation}
   E_s \propto \frac{\phi_s}{(\dw-l)} \; \; \; \; \; \;\; \;  (l < \dw)
\end{equation}
As $l \rightarrow \dw$,  $E_s$ is very  large at the tips, leading
to stringy patterns and  the  minima in Fig.6.  In a  recent  work
\cite{john95a}, it has  been shown that $\Delta  W$ and $l$ can be
correlated to the depletion layer width and barrier lowering given
in equations (1) and (2).

 The existence of  competing  processes in electrochemical deposition
and other phenomena  is well established. The phase diagrams reported
by  Sawada  {\it et al.}  and  Grier {\it  et al.} represents various
morphological classes resulting from a variation of the concentration
and  applied  voltage. Assigning  a direct  correlation  between  our
parameters and the experimental  parameters based  on  a comparison
between the  phase diagrams  of the simulation  and  experiment is
difficult. In all probability,  $\Delta W$ and  $l$ are  functions of
both concentration and applied  voltage.  Further  work  needs to  be
done  to  establish  the exact functional correlations between $\Delta W$
and $l$ and the physical processes involved in electrochemical
deposition.

	In a sense, the two scales \dw and $l$ are  antithetical, one
separating the  particle  from the  aggregate, the  other  driving it
towards  the   aggregate.  Various  growth   models  (DLA,  ballistic
deposition,  Eden)  can  be  obtained  as   limiting  cases  of  this
algorithm. We stress that the  two scales do not contain  any {\em  a
priori}   bias  towards   any   specific   morphological   structure.
Nevertheless, a variety of distinct morphological structures observed
in experimental growth and dissolution is obtained in our simulation.

\acknowledgements
	Useful discussions with Dr.D.  Dhar and Dr.D.Chowdhury are
gratefully acknowledged. We thank  Prof.B.M.Arora, SSE group, TIFR
where part   of  the work   was  carried out.   One    of us (GCJ)
acknowledge financial support from the  Council for Scientific and
Industrial Research, Government of India.


\begin{thebibliography}{10}

\bibitem{benj90}
E. Ben-Jacob and P.Garik, Nature {\bf 343},  523  (1990).

\bibitem{smit92}
R.L.Smith and S.D.Collins, J. Appl. Phys. {\bf 71},  R1  (1992).

\bibitem{sawa86}
Y.Sawada, A.Dougherty, and J.P.Gollub, Phys. Rev. Lett. {\bf 56},  1260
  (1986).

\bibitem{grie86}
D.Grier, E.Ben-Jacob, R. Clarke, and L.M.Sander, Phys. Rev. Lett. {\bf 56},
  1264  (1986).

\bibitem{beal85}
M.I.J.Beale {\it et~al.}, J. Cryst. Growth {\bf 73},  622  (1985).

\bibitem{mull63}
W.W.Mullins and R.F.Sekerka, J. Appl. Phys. {\bf 34},  323  (1963).

\bibitem{grie87}
D.~G. Grier, D.~A. Kessler, and L.M.Sander, Phys. Rev. Lett. {\bf 59},  2315
  (1987).

\bibitem{gari89}
P.Garik {\it et~al.}, Phys. Rev. Lett. {\bf 62},  2703  (1989).

\bibitem{herr86}
H.J.Herrmann, Phys. Rep. {\bf 136},  153  (1986).

\bibitem{meak88}
P.Meakin,  in {\em Phase Transitions and Critical Phenomena Vol.12}, edited by
  C.Domb and T.L.Lebowitz (Academic, New York, 1988).

\bibitem{stan85}
{\em On Growth and Form}, {\em NATO ASI Series E No 100}, edited by H.E.Stanley
  and N.Ostrowsky (Martinus Nijhoff, Cargese, France, 1985).

\bibitem{witt81}
T.A.Witten and L.M.Sander, Phys. Rev. Lett. {\bf 47},  1400  (1981).

\bibitem{vics84}
T. Vicsek, Phys. Rev. Lett. {\bf 53},  2281  (1984).

\bibitem{erle93}
J.Erlebacher, P.C.Searson, and K.Sieradzki, Phys. Rev. Lett. {\bf 71},  3311
  (1993).

\bibitem{shoc92a}
O.Shochet {\it et~al.}, Physica A {\bf 181},  136  (1992).

\bibitem{shoc92b}
O.Shochet {\it et~al.}, Physica A {\bf 187},  87  (1992).

\bibitem{smit88}
R.L.Smith, S.-F.Chuang, and S.D.Collins, J. Electron. Mater. {\bf 17},  533
  (1988).

\bibitem{smit89}
R.L.Smith and S.D.Collins, Phys. Rev. A {\bf 39},  5409  (1989).

\bibitem{voss84}
R.F.Voss, J. Stat. Phys. {\bf 36},  861  (1984).

\bibitem{ball85}
R.C.Ball and R.M.Brady, J.Phys.A {\bf 18},  L809  (1985).

\bibitem{erle94}
J.Erlebacher, K.Sieradzki, and P.C.Searson, J. Appl. Phys. {\bf 76},  182
  (1994).

\bibitem{john95a}
G.C.John and V.A.Singh, Phys. Rev. B {\bf 52},  11125  (1995).

\bibitem{benj86}
E.Ben-Jacob {\it et~al.}, Phys. Rev. Lett. {\bf 57},  1903  (1986).

\bibitem{phase}
The term ``phases'' is taken to mean classes of aggregates as reported earier
  in literature.

\bibitem{mats85}
M.Matsushita, Y.Hayakawa, and Y.Sawada, Phys. Rev. A {\bf 32},  3814  (1985).

\bibitem{meak83}
P. Meakin, Phys. Rev. A {\bf 27},  2616  (1983).

\bibitem{bowe91}
R.W.Bower and S.D.Collins, Phys. Rev. A {\bf 43},  3165  (1991).

\end{thebibliography}

\begin{figure}
\caption{The drift-diffusion model. A particle is released beyond
the depletion layer boundary (the closed curve) which dynamically
follows the aggregate (solid circle cluster) contour at a distance
$\Delta W$. When the particle wanders to within a radius $l$ (see
dotted arrow) of the aggregate surface, it is ``field driven'' to
become a part of the cluster.}
\end{figure}

\begin{figure}
\caption{Map of the simulated growth patterns with varying depletion
width `\dw'  and  drift length  `$l$'.  (a)  Patterns  resembling DLA
($\dw=10$,  $l=2$),  (b) stringy structures  ($\dw=10$, $l=10$),  (c)
homogeneous,  dense  branching morphologies  with a  nearly  circular
growth front  ($\dw=2$,  $l=2$) and  (d) thick  dendritic growth with
side branches ($\dw=8$ and $l=24$).}
\end{figure}
\begin{figure}
\caption{The fractal dimension $d_f$ plotted against $\Delta W$
(in lattice units)
with  $l=2$. For  small $\Delta    W$,  $d_f\simeq 2$, which    is
indicative of  compact clusters.  In  the large  $\Delta W$ limit,
$d_f$ stabilizes at $1.65\pm 0.05$.}
\end{figure}
\begin{figure}
\caption{The fractal dimension $d_f$ plotted against $\Delta W$
(in lattice units)
with $l$ equal to $\Delta W$. $d_f$ is seen to approach unity in
the large $\Delta W (=l)$ limit.}
\end{figure}
\begin{figure}
\caption{``Phase'' plot of patterns observed in the simulations. The
phases depicted transform continuously into one another. Hence, the
boundaries must not be treated as rigidly defining a transition.
Both $\Delta W$ and $l$ are expressed in terms of lattice units.}
\end{figure}
\begin{figure}
\caption{Plot of the percentage aggregate density vs. $l$ (in lattice units)
obtained in deposition
on to a linear substrate of  length 300 units.  The two curves are
for  $\dw=6$ and $\dw=9$. In  each case, a  minimum is observed at
$\dw=l$.}
\end{figure}
\begin{figure}
\caption{(a).  Percentage aggregate density vs depth (in lattice
units)  plot  for  the  deposition  on  to   a   linear substrate.
Simulation done for a 500$\times$500 lattice with $\Delta W=5$ and
$l=2$.  A cross-over to a  non-fractal behavior where the  density
is almost depth independent is seen.  (b) Log-log plot of the mass
vs cluster radius for radial deposition  on to a central seed. The
clusters are seen to   be scale invariant  over  a large  range of
diameters (100-1000 lattice units).}
\end{figure}
\end{document}